\documentclass{article}
\usepackage{PRIMEarxiv}

\usepackage[utf8]{inputenc} % allow utf-8 input
\usepackage[T1]{fontenc}    % use 8-bit T1 fonts
\usepackage{hyperref}       % hyperlinks
\usepackage{url}            % simple URL typesetting
\usepackage{booktabs}       % professional-quality tables
\usepackage{amsfonts}       % blackboard math symbols
\usepackage{nicefrac}       % compact symbols for 1/2, etc.
\usepackage{microtype}      % microtypography
\usepackage{xcolor}         % colors
\usepackage{multirow}
\usepackage{multicol}
\usepackage{amsfonts}       % blackboard math symbols
\usepackage{graphicx}

\usepackage{fancyvrb}
\usepackage{xspace}
\usepackage{listings}
\usepackage{soul}
\usepackage{adjustbox}

\newcommand{\verify}{\texttt{VerifyThisBench}\xspace}
\newcommand{\verirelax}{\texttt{VerifyThisBenchXS}\xspace}

\definecolor{codegreen}{rgb}{0,0.6,0}
\definecolor{codepurple}{rgb}{0.58,0,0.82}
\definecolor{codeyellow}{rgb}{1,0.96,0.7}
\definecolor{codegray}{rgb}{0.5,0.5,0.5}
\definecolor{lightcoral}{rgb}{1,0.9,0.8}
\definecolor{backcolour}{rgb}{0.95,0.95,0.92}

\newcommand{\citet}[1]{(\cite{#1})}
\newcommand{\citep}[1]{(\cite{#1})}
\newcommand{\iclr}[1]{\textcolor{black}{#1}}

\title{
VerifyThisBench: Generating Code, Specifications, and Proofs All at Once}

\author{
  Xun Deng \\
  University of Toronto\\
  \texttt{xun.deng@mail.utoronto.ca} \\
   \And
  Sicheng Zhong \\
  University of Toronto \\  \texttt{sicheng.zhong@mail.utoronto.ca} \\
  \And
  Barış Bayazıt\\
  University of Toronto \\
  \texttt{baris@cs.toronto.edu}\\
  \And
  Andreas Veneris \\
  Univerisity of Toronto \\
  \texttt{veneris@eecg.toronto.edu} \\
  \And
  Fan Long \\
  University of Toronto \\
  \texttt{fanl@cs.toronto.edu} \\
   \And
  Xujie Si \\
  University of Toronto \& Vector Institute\\
  \texttt{six@cs.toronto.edu} \\
}

\begin{document}
\maketitle
\begin{abstract}

Large language models (LLMs) have demonstrated remarkable progress in code generation, but many existing benchmarks are approaching saturation and offer little guarantee on the trustworthiness of the generated programs.
To improve visibility into model reasoning on formal correctness, we introduce \verify, 
%% a new benchmark designed to evaluate LLMs on end-to-end program verification tasks that require interpreting natural language problem descriptions, formulating formal specifications, generating code, and constructing correctness proofs.
a new benchmark that evaluates end‑to‑end program verification from natural language descriptions: models must (i) extract formal specifications, (ii) implement in a verification‑aware language, and (iii) construct machine‑checkable proofs.
Our evaluation reveals that even state-of-the-art (SOTA) models, such as o3-mini, achieve a pass rate of less than 4\%, with many outputs failing to compile. To isolate sources of difficulty, we further propose \verirelax, a relaxed variant in which partial implementations or proofs are provided. 
%We systematically assess SOTA models on both benchmarks, uncovering key strengths and limitations in their formal reasoning and verification capabilities.
Across nine models and seven verification tools on both benchmarks, we observe consistent gains with feedback‑driven refinement, but overall pass rates remain low, underscoring substantial gaps in formal reasoning. We release the benchmark and the unified evaluation environment to catalyze the verification capabilities for future models.

\end{abstract}

% keywords can be removed

\section{Introduction}

Large language models (LLMs) have unequivocally revolutionized the landscape of automated code generation. Models like \citet{openai2024gpt4o} GPT-4o, \citet{google2024gemini} Gemini, \citet{claude2024introducingV3} Claude, and \citet{copilot2021introducing} Copilot excel at generating functional code snippets and translating between languages. These capabilities are now integrated into AI-powered IDEs, such as Cursor and Visual Studio, to support large-scale software development. This proficiency had led to increasing needs on established benchmarks, as early as HumanEval~\cite{chen2021evaluatinglargelanguagemodels} and MBPP~\cite{austin2021programsynthesislargelanguage}, to reflect the capability of each LLM tool. However, this rapid progress raises critical questions about the trustworthiness and reliability of the generated artifacts. Many existing benchmarks, while useful for gauging functional correctness through test suites, are approaching saturation~\citep{DynaBench21,AILuminate25,CRUXEval24,xia2024top} and inherently offer limited guarantees regarding the deeper aspects of program correctness. Test cases, by their nature, can demonstrate the presence of bugs, but cannot prove their absence~\cite{dijkstra1972humble}, leaving a significant gap in assessing the formal robustness and true reasoning capabilities of these powerful models.

Reliable software must go beyond passing tests to be trustworthy, precisely follow specifications, and even self-validate. Formal verification offers the most rigorous approach to achieving these guarantees. This paradigm involves providing machine-checked mathematical proofs to show that a program adheres to its formal specification, thereby guaranteeing critical properties such as functional correctness, liveness (ensuring the program eventually does something good), and safety (ensuring the program never does something bad)~\cite{huth2004logic}. Modern program verification infrastructures, such as Dafny~\cite{leino2010dafny}, Frama-C~\cite{kirchner2015frama}, Verus~\cite{lattuada2023verus}, Isabelle/HOL~\cite{nipkow2002isabelle}, and Lean~\cite{deMoura2018lean}, coupled with powerful automated theorem provers and SMT solvers like Z3~\cite{demoura2008z3} and CVC5~\cite{barbosa2022cvc5}, have significantly streamlined the process of writing and checking such verified software. These tools allow developers to express complex specifications and then automatically or semi-automatically verify that the implementation meets these specifications.

Although researchers have developed multiple benchmarks to assess LLMs on formal verification subtasks~\cite{kamath2023finding,rankInv23,LLMinv23,LLMpostcondition24}, none evaluates end-to-end program verification solely from natural-language inputs. Instead, existing suites either require verifying or synthesizing small programs against a given formal specification, or focus on aiding proof completion by suggesting individual verification steps. Consequently, even though state-of-the-art LLMs have been reported to solve up to 97.8\% of these benchmark tasks~\cite{LLMinv-ASE24}, those numbers do not reflect their true capability for end-to-end program verification.

%Despite the availability of these powerful verification frameworks, their integration into the LLM-driven code generation pipeline remains nascent. 

% \comment{add advantages of using verifythis challenge to differentiate from parallel work. }
% \xun{pls help discuss on this, as there are also parellel work bridging this gap.}

To bridge this gap and rigorously evaluate the capabilities of LLMs in this demanding domain, we introduce \verify, a novel benchmark designed to assess end-to-end program verification, as shown in Figure~\ref{fig:overview}. Inspired by the annual \citet{verifythis} where human contestants devise implementations and accompanying formal proofs in verification-aware languages, \verify tasks LLMs with interpreting natural language problem descriptions, formulating formal specifications, generating the corresponding code, and constructing machine-checkable correctness proofs -- all at once, to produce compiled and verified artifacts. \iclr{While recent efforts ~\cite{ye2025verina,thakur2025clevercuratedbenchmarkformally} also benchmark LLMs on end-to-end verification tasks in Lean, our work differs by building on the long-standing VerifyThis Challenge, offering multi-framework coverage, research-grade tasks, and competition-vetted difficulty, with solution lengths up to 648 lines compared to a maximum of 225 in prior work.}

Our evaluation using \verify reveals that even state-of-the-art (SOTA) models, such as o3-mini, achieve a zero-shot pass rate of 3.62\% on this end-to-end task, with a significant number of outputs failing even to compile, and only reach a pass rate of 9.37\% after five rounds of feedback.
These results underscore the profound challenge this domain presents. To dissect these challenges further and explore capabilities in a more guided setting, we also propose \verirelax, a variant where partial specification, implementation code, or proofs are provided, and the LLM is tasked to complete the missing components. \iclr{In this setting, o3-mini achieves 2.24\% in zero-shot attempt and 8.28\% after refinement.}

% add numbers for verifythisbenchXS

This paper makes the following key contributions:
\begin{itemize}
\item \textbf{VerifyThisBench:} We present \verify, a new benchmark suite for evaluating the ability of LLMs to generate fully verified programs (code, specifications, and proofs) from natural language descriptions.

\item \textbf{Relaxed VerifyThisBench:} We introduce \verirelax, a relaxed version of the \verify, to assess LLM performance when provided with partial artifacts and tasked with completing them.

\item \textbf{Unified Environment:} We provide a unified evaluation environment that integrates seven verification tools and an automated pipeline, enabling consistent and scalable benchmarking across diverse formal verification tasks.

\item \textbf{SOTA LLM Evaluation:} We conduct a systematic evaluation of nine SOTA LLMs on both benchmarks, revealing current capabilities and significant limitations. 

%Our analysis includes performance breakdowns across tools, attempt-based comparisons, model-specific strengths, self-assessed coherence, and the impact of partial guidance, providing a comprehensive understanding of model behavior in formal verification tasks.

\begin{figure}
    \centering
\includegraphics[width=0.9\linewidth]{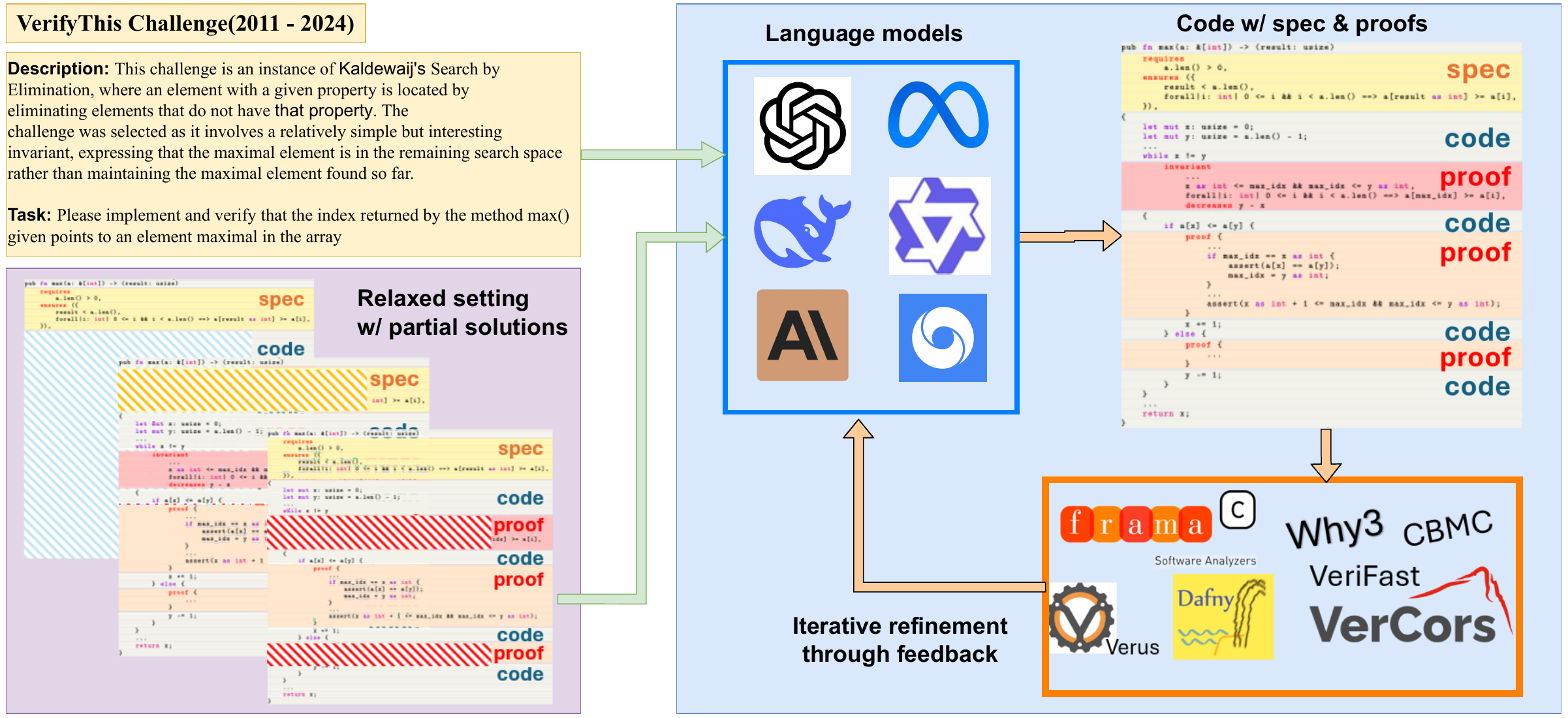}
    \caption{Evaluation workflow of \verify and its relaxed settings.}
    \label{fig:overview}
\end{figure}

% \item \textbf{Ablation Studies: } Through a series of ablation studies, we analyze the impact of different prompting strategies (direct generation, iterative refinement with compiler feedback) on model performance.
% \item \textbf{Failure Mode Analysis: } We provide an in-depth analysis of common failure modes, offering insights into the specific challenges LLMs face in formal specification synthesis, code generation tailored for verification, and automated proof construction.
\end{itemize}

\section{Background \& Related Work}

% \subsection{code-related benchmark}
% Human-Eval has been a long studied benchmark in the field of code generation. Its has been extended other languages and benchamrk in similar forms has been proposed. However, SOTA models have begun to nearly saturate these benchmarks, reaching near perfect performance. This suggests that these benchmarks now offer a limited discrimination among strong models. 
% SWEBench is a software engineering tasks grounded in real-work github issues. However, many of these task are impossible to solve, thus does not fully capture the performance of the large language model. In this work, we propose a challenging benchmark consiting of program verification tasks, evaluating the program using formal methods. 

% \subsection{program verification benchmark}
% Code2Inv (very small programs, 5~10 lines)

% SV-COMP (large programs, legacy code, no code generation, only verification)

\subsection{Unverified Code Synthesis Benchmarks}
Recent benchmarks for code generation include \textbf{APPS}~\cite{hendrycks2021measuringcodingchallengecompetence}, \textbf{HumanEval}~\cite{chen2021evaluatinglargelanguagemodels}, \textbf{MBPP}~\cite{austin2021programsynthesislargelanguage}, \textbf{CodeContests}~\cite{Li_2022}, \textbf{DS-1000}\cite{lai2022ds1000naturalreliablebenchmark}, \textbf{SWEBench}~\cite{jimenez2024swebenchlanguagemodelsresolve}, and \textbf{EvalPlus}~\cite{liu2023codegeneratedchatgptreally}, among others. These benchmarks present programming tasks, often sourced from online competitions or community platforms, and evaluate models based on whether generated solutions pass a set of input-output test cases. While effective in emulating daily software development, they do not involve formal verification.

In contrast, \verify requires models to go beyond functional testing: they must formalize natural language intents into specifications, generate code in verification-aware languages, and produce proofs that pass a formal logic verifier. This makes \verify a substantially more rigorous and comprehensive benchmark than traditional code synthesis tasks.

\subsection{Program Verification Benchmarks}
Benchmarks built in the context of formal verification include \textbf{SV-COMP}~\cite{SVCOMP}, \textbf{SyGuS}~\cite{sygus}, and \textbf{Code2Inv}~\cite{si2020code2inv}. \textbf{SV-COMP} and \textbf{Code2Inv} focus solely on verification tasks that do not require implementation generation. For more contexts, the former contains large-scale C/Java benchmarks verifying fixed safety properties, and the latter targets loop-invariant generation over small C-style programs. \textbf{SyGuS} focuses on constraint-based synthesis. 

More recent efforts like \textbf{DafnyBench} ~\cite{loughridge2024dafnyBench} and \textbf{VerusBench} ~\cite{yang2024verusBench} collect verified programs in Dafny and Verus respectively, primarily to train and evaluate ML-based tools in aiding proof completion and suggesting verification steps, rather than end-to-end program generation from natural language.

These benchmarks evaluate components of the verification pipeline but typically assume a preset formal specification or verification goal. In contrast, \verify %evaluates the full pipeline
%: starting from a natural-language task, generating in formal language to satisfy a formal verifier. 
uses the end-to-end setup to explicitly evaluate the model's ability in interpreting and encoding natural-language descriptions into provably correct formal programs, a capability not tested in existing benchmarks.

\subsection{End-to-end Verification Benchmarks}
\iclr{Parallel work includes \textbf{Verina}~\cite{ye2025verina} (189 tasks) and \textbf{Clever}~\cite{thakur2025clevercuratedbenchmarkformally} (161 tasks), exploring end-to-end verification in Lean, with sources translated from programming tasks in \textbf{HumanEval}, \textbf{MBPP} and \textbf{Leetcode }etc. \verify differs in the source, scope and diversity, with 734 tasks derived from the VerifyThis competition series, which presents realistic, research-grade verification challenges across multiple domains. Rather than focusing on Lean, \verify spans seven verification frameworks across multiple programming languages, including Verus~\cite{lattuada2023verus} and Frama-C~\cite{kirchner2015frama} that are established for production codebases. Moreover, \verify includes tasks that require reasoning about memory safety, concurrency, and complex data structures beyond arrays and trees. See Appendix~\ref{app:loc}  for detailed comparison of description lengths and solution sizes.}%\xun{check if the above is a good discussion and delete the rest}
%verifying existing Rust and C codebases within production environments

% \comment{The stats for verina and clever is available as well. Maybe we should compare them here, or a table in appendix} \xun {maybe we can not provide stats here but discuss in the appendix}Our benchmark is also broader in difficulty: natural language descriptions range from 30 to 1802 words, and reference implementations vary from as short as 28 lines to nearly 700 lines, with a median of 124 lines. 

% add more two sentences

\subsection{Formal Methods in Software Verification: A Primer}

Formal methods in software verification aim to mathematically prove program correctness against a \textbf{formal specification}—a precise, unambiguous description of what a program should do, often expressed in a logical language. This contrasts with testing, which can only show the presence of bugs for specific inputs. The verification process typically relies on several key components embedded within or alongside the executable program code:
\begin{itemize}
    \item Contracts: These formalize the obligations and guarantees of a code segment.
    \begin{itemize}
        \item Pre-conditions (\texttt{requires} clauses): Properties that must hold true before a function or code block executes for it to behave correctly.
        \item Post-conditions (\texttt{ensures} clauses): Properties guaranteed to be true after a function or code block finishes, provided its pre-conditions were met.
    \end{itemize}
    \item Intermediate Assertions: Assistive hints are often needed to bridge any reasoning gaps between the pre\&post-conditions where the underlying solver cannot automatically address.
    \item Loop Invariants: For iterative constructs, loop invariants are crucial properties that hold at the start of a loop, are preserved by each iteration, and, in conjunction with the loop's termination, help prove the loop's correctness.
\end{itemize}

% Formal methods in software engineering are mathematically rigorous techniques used to specify, develop, and verify software and hardware systems \cite{huth2004logic}. Unlike empirical testing, which can only show the presence of bugs for specific inputs, formal verification aims to prove or disprove the correctness of a program with respect to a certain formal specification or property for all possible inputs. This process typically involves several key components:

% \textbf{Formal Specification} This is the cornerstone of formal verification. A formal specification is a precise, unambiguous description of what a program (or a part of it, like a function or module) is intended to do. Specifications are written in a language with a formally defined syntax and semantics, often based on mathematical logic (e.g., first-order logic, temporal logic). For instance, a specification for a sorting function would formally state that for any given input array, the output array must be a permutation of the input and its elements must be in non-decreasing order.

The typical verification flow in systems utilizing these concepts is as follows:
\begin{enumerate}
    \item \textbf{Annotation:} Developers write code in a verification-aware language (e.g., Dafny~\cite{leino2010dafny}, Frama-C, Verus and annotate it with formal specifications and proof hints, including pre-conditions, post-conditions, assertions, and loop invariants.
    \item \textbf{Generation of Proof Obligations:} A tool, often a Verification Condition Generator (VCG), processes the annotated code and its specifications. It translates them into a series of mathematical proof obligations (verification conditions) that, if all true, logically imply the program's correctness with respect to its specification.
    \item \textbf{Automated Proving:} These verification conditions are then fed to backend automated theorem provers, typically Satisfiability Modulo Theories (SMT) solvers like Z3~\cite{demoura2008z3} or CVC5~\cite{barbosa2022cvc5}. These solvers attempt to mathematically prove each obligation.
    \item \textbf{Feedback:} The system reports to the developer whether the proofs succeeded or failed. Failures often pinpoint inconsistencies between the code and its specification, or missing/incorrect annotations.
\end{enumerate}

Successfully generating code within this paradigm, as targeted by our \verify, requires an LLM not only to produce the algorithmic implementation but also to understand, formulate, and correctly express intricate formal specifications and proof structures that enable automated verification.

% Formal methods 
% - specification
% - pre/post condition
% - invariants

\section{VerifyThisBench Benchmark}
\verify~is inspired by the annual \textbf{VerifyThis Challenges}~\cite{verifythis}, a competition where participants are tasked with formalizing specifications, implementing solutions, and verifying that the implementations meet the specification. \iclr{We focus on this benchmark because it is a dedicated formal methods competition, designed not only to evaluate participants’ skills but also to assess the maturity of verification tools~\cite{10.1007/978-3-031-67695-6_5}.}
Each challenge is designed to be completed within a 90-minute session and varies in difficulty. Submissions are evaluated based on correctness, completeness, and additional quality criteria such as elegance and the degree of automation. Similarly, in \verify, the task is to interpret natural language problem descriptions and implement code and write proofs. 

% Every task in \verify~begins from an informal description. Problems are stated in natural language. The model must interpret this description and formalize the underlying requirements by writing appropriate specifications (preconditions, postconditions, invariants, etc.). In other words, \verify~tests the ability of an LLM to translate under-specified, human-readable problem descriptions into precise formal logic. This contrasts with benchmarks that supply already-formal inputs; here the model must infer the intended specification from context.

\subsection{Benchmark Construction} 
We collected challenges from the annual competition series between 2011 and 2024, each with natural-language descriptions (seldom include pseudo-code) and associated (one or more) tasks.
%documenting their descriptions, pseudo-code, and associated tasks. 
Tasks are categorized as either implementation (completing an algorithm) or verification (proving a model or implementation correct against a specification).
%, all described in natural language only. 
The resulting dataset includes 41 challenges and 154 tasks, with an example in Appendix \ref{app:example} and detailed statistics in Appendix \ref{app:loc}. 
\iclr{The dataset is available in supplementary material and will be made public after the anonymity period.}

%The dataset is available at~\cite{verifythisbench}. 

\subsection{Environment} 
To facilitate evaluation, we provide a unified environment supporting seven verification tools. Five of them, \textbf{Dafny}~\cite{leino2010dafny}, \textbf{\citet{why3}}, \textbf{\citet{VerifastGithub}}, \textbf{\citet{VercorsGithub}}, %\textbf{VeriFast}~, \textbf{VerCors}~, 
and \textbf{Frama-C}~\cite{kirchner2015frama}, are widely used in past VerifyThis competitions. To broaden tool diversity, we additionally include \textbf{Verus}~\cite{lattuada2023verus} and \textbf{CBMC}~\cite{kroening2023cbmccboundedmodel}, covering Rust, C, and other imperative or deductive platforms. Tool versions and brief descriptions can be found in Appendix~\ref{toolversion}. 

\subsection{Features of VerifyThisBench}
\textbf{End-to-end verification tasks with natural language problem descriptions}: All tasks start with informal, natural language prompts (often with pseudo-code). Models must interpret the intent and formalize it into precise logical specifications. They are required to generate specifications, implementations, and formal proofs in a verification-aware language, ensuring the code passes machine-checkable verification. Example challenge and solution can be found in Appendix~\ref{app:example}.

\textbf{Graded difficulty and multi-step challenges:} Challenges are drawn from the VerifyThis competition and span a range of difficulties (see Appendix~\ref{app:loc}.). Many include sequential subtasks, allowing fine-grained assessment of model capability on step-wise tasks.
 
\textbf{Tool diversity:}
Multiple tools are provided and tested on. Models must conform to the syntax and semantics of real-world verification frameworks.

\subsection{Relaxation}
We observe that most language models fail to generate compilable code when targeting specific verification tools. This is often due to the syntactic complexity and precise annotations required by these tools. To isolate the sources of difficulty and better assess LLM capabilities under more supportive conditions, we construct a set of relaxed subtasks derived from past human-written solutions. Specifically, we define three forms of relaxation. In \textbf{Code-Gen}, we provide the function specifications, omitting both the implementation and the proof annotations.
In \textbf{Specification-Gen}, we provide the implementation and its proof, but remove the function specifications.
In \textbf{Loop-Gen}, we provide specifications and implementations, but remove loop invariants needed for verification.

% To further diversify the difficulty spectrum, we vary the extent of relaxation. In some instances, we remove all relevant components (e.g., entire specs or proofs), while in others, we retain partial elements or include complete examples as guidance. This enables a more graded evaluation of LLM performance across varying levels of verification support.
In total, we create a set of 580 tasks. Specifically, there are 226 code-gen task, 121 loop-gen tasks, and 233 spec-gen tasks. Table~\ref{tab:RelaxStats} in Appendix~\ref{relaxbytool} shows the statistics of \verirelax. \iclr{Since no prior solutions exist for CBMC and Verus, and given notable community interest, we developed new Verus solutions to enrich the dataset; CBMC solutions remain unavailable and are therefore not included in the relaxed experiments.}

% \textbf{Relaxed Variant:} Models can be evaluated in a more nuanced way. By comparing model performance on the original problem versus its relaxed variant, researchers can assess how much assistance or simplification is needed for success. This design explicitly enables the study of in-context learning and graduated verification support: for example, one can test how providing a hint or partial proof in the prompt improves the outcome. In short, the \verirelax\ tasks allow a fine-grained analysis of formal reasoning capability under varying levels of difficulty and guidance.}

\section{Experiment Results}
\subsection{Model Setup}

We evaluate a diverse set of SOTA language models, covering both proprietary and open-source systems. Representatives are selected from the \citet{OpenAI} family (GPT-4o, GPT-4omini, o3-mini, o4-mini), \citet{claude37} (Claude-3.7-Sonnet), Google (Gemini-2.5-Flash)~\cite{gemini}, DeepSeek (Deepseek-chat-v3)~\cite{deepseekai2024deepseekv3technicalreport}, \citet{metallama} (Llama3.3-70B-Instruct) and Alibaba (Qwen-2.5-72B-Instruct)~\cite{qwen2.5}. This selection enables a comprehensive comparison across different model architectures and training paradigms. Model versions are provided in Appendix~\ref{version}.

\subsection{Experiment Design and Metrics}

For both \verify~and \verirelax, we conduct experiments with iterative refinement based on tool-generated error messages. To evaluate correctness, we pass the generated code to the target verification tool and check whether it compiles and verifies successfully. A task is marked as pass/succeed if no error is returned.

In addition to correctness checking, we introduce a coherence check as a relaxed evaluation metric. In this step, the model self-assesses whether its generated code semantically aligns with the original problem intent -- an aspect difficult to verify automatically. This metric helps determine how well the specification matches the task description and provides insight into the model’s ability in auto‑formalization and symbol grounding.

Each task is attempted five times per model. The first attempt uses only the task prompt; the next four incorporate feedback from previous errors. During refinement, the model has access to the full history of its prior attempts and corresponding feedback for the current task, enabling iterative correction.

In \verify, a challenge may have multi-stage tasks that are completed sequentially. Only the final attempt from the previous subtask is carried over to the next, preserving essential contexts while keeping the prompts concise. In contrast, \verirelax tasks have isolated contexts and are completed independently, with no progress carried over between tasks.

To ensure fairness, we use the same prompt (see Appendix~\ref{app:prompt}) across all models and set the temperature to 0.7 when applicable. Timeout of one minute is enforced for all experiments on the verifiers. The experiments were conducted on a machine with an Intel i7-1360P CPU and 16GB of RAM. 

\subsection{Overall Pass Rate}
Table~\ref{tab:verifyoverall} presents the performance of the SOTA models on \verify. For each verification tool, we report pass rates on the initial zero-shot attempt and after four additional refinement attempts using feedback.

In the first attempt, most models perform poorly, with success rates under 4\%. The top performers are o3-mini, Llama, and Claude, indicating that even the strongest models struggle initially. By the fifth attempt, performance improves significantly across all models. o3-mini leads overall, followed by Claude, o4-mini, and Llama. These results highlight the effectiveness of iterative refinement and feedback in enhancing model performance.

Each model exhibits distinct strengths across different verification tools, underscoring that no single model consistently outperforms the rest. For example, o3-mini, the top overall performer, excels especially in CBMC and Verus. On the other hand, Claude shows consistent strength in Dafny and Frama-C.  Gemini, while generally average, performs exceptionally well on VerCors. Llama, another open-source model, performs best on Verus. In contrast, Qwen shows consistently low performance across all tools, suggesting limitations in its current proof synthesis capabilities. Further insights into tool-specific performance are discussed in Section~\ref{sec:tool}.

\begin{table}[!htp]\centering
\caption{Overall Pass Rate On \verify}\label{tab:verifyoverall}
\scriptsize
\begin{adjustbox}{width=0.9\textwidth} 
\begin{tabular}{lrrrrrrrrrrr}\toprule
& Attempt &GPT4o &GPT4o-mini &o3-mini &o4-mini &Claude &Gemini &Llama &Deepseek &Qwen \\\midrule
CBMC &zero-shot &\textbf{8.44\%}&7.14\% &\textbf{8.44\%} &1.30\% &6.49\% &1.95\% &7.14\% &0.65\% &1.30\% \\
&refinement &20.13\% &19.48\% &\textbf{25.32\%} &15.58\% &22.08\% &14.94\% &20.13\% &22.08\% &3.25\% \\
\midrule
Dafny &zero-shot &0 &0 &\textbf{4.55\%} &1.95\% &3.90\% &0 &0 &1.30\% &0 \\
&refinement &1.30\% &0.65\% &9.74\% &10.39\% &\textbf{11.04\%} &1.30\% &2.60\% &2.60\% &0.65\% \\
\midrule
Frama-C &zero-shot &0 &0.65\% &0 &0 &\textbf{3.90\%}&0.65\% &0 &1.95\% &0 \\
&refinement &7.14\% &1.95\% &2.60\% &3.25\% &\textbf{11.04\%} &8.44\% &0.65\% &3.25\% &0 \\
\midrule
VerCors &zero-shot &0 &1.30\% &1.30\% &1.95\% &0 &5.84\% &\textbf{8.44\%} &1.95\% &0 \\
&refinement &1.95\% &1.30\% &1.95\% &5.19\% &1.30\% &\textbf{16.88\%} &11.69\% &4.55\% &3.25\% \\
\midrule
VeriFast &zero-shot &0 &0 &0 &0 &0 &0 &0 &0 &0 \\
&refinement &0 &0 &0 &\textbf{2.60\%} &0 &0 &0.65\% &0.65\% &0 \\
\midrule
Verus &zero-shot &1.95\% &6.49\% &\textbf{10.39\%} &0.65\% &0.65\% &0.65\% &7.79\% &0.65\% &0.65\% \\
&refinement &12.99\% &9.09\% &\textbf{21.43\%} &8.44\% &1.30\% &0.65\% &17.53\% &1.30\% &0.65\% \\
\midrule
Why3 &zero-shot &0 &0 &0.65\% &0.65\% &\textbf{1.30\%} &\textbf{1.30\%} &0 &0.65\% &0 \\
&refinement &0 &0 &4.55\% &\textbf{10.39\%} &9.09\% &5.84\% &1.95\% &1.95\% &0 \\
\midrule
Overall &zero-shot &1.48\% &2.23\% &\textbf{3.62\%} &0.93\% &2.32\% &1.48\% &3.34\% &1.02\% &0.28\% \\
&refinement &6.22\% &4.64\% &\textbf{9.37\%} &7.98\% &7.98\% &6.86\% &7.88\% &5.19\% &1.11\% \\
\midrule
Improvement & &4.73\% &2.41\% &5.75\% &\textbf{7.05\%} &5.66\% &5.38\% &4.55\% &4.17\% &0.83\% \\
\bottomrule
\end{tabular}
\end{adjustbox}
\end{table}

\begin{table}[!htp]\centering
\caption{Overall Pass Rate On \verirelax}\label{tab:relaxoverall}
\scriptsize
\begin{adjustbox}{width=0.9\textwidth} 
\begin{tabular}{lrrrrrrrrrrr}\toprule
& Attempt &GPT4o &GPT4o-mini &o3-mini &o4-mini &Claude &Gemini &Llama &Deepseek &Qwen \\\midrule
% cbmc &zero-shot & & & & & & & & & \\
% &refinement & & & & & & & & & \\
Dafny &zero-shot &0 &1.35\% &\textbf{2.70\%} &1.35\% &1.35\% &1.35\% &0 &\textbf{2.70\%} &1.35\% \\
&refinement &17.57\% &9.46\% &31.08\% &37.84\% &\textbf{41.89\%} &17.57\% &8.11\% &21.62\% &8.11\% \\
\midrule
Frama-C &zero-shot &0 &0 &1.85\% &0 &1.85\% &1.85\% &0 &1.85\% &0 \\
&refinement &0 &0 &5.56\% &\textbf{18.52\%} &9.26\% &1.85\% &0 &5.56\% &0 \\
\midrule
VerCors &zero-shot &0 &0 &0 &0 &0 &0 &0 &0 &0 \\
&refinement &0 &0 &0 &\textbf{7.69\%} &0 &0 &3.85\% &0 &0 \\
\midrule
VeriFast &zero-shot &7.58\% &4.55\% &3.03\% &6.06\% &6.06\% &0 &4.55\% &\textbf{12.12\%} &3.03\% \\
&refinement &12.12\% &6.06\% &4.55\% &10.61\% &\textbf{27.27\%} &0 &9.09\% &13.64\% &6.06\% \\\midrule

Verus &zero-shot &7.07\% &5.05\% &8.08\% &14.14\% &14.14\% &4.04\% &3.03\% &7.07\% &5.05\% \\

&refinement &15.15\% &6.06\% &17.17\% &30.30\% &30.30\% &16.16\% &13.13\% &20.20\% &7.07\% \\
\midrule
Why3 &zero-shot &0 &0 &0 &0.38\% &0.38\% &\textbf{0.77\%} &
0 &0.38\% &0 \\
&refinement &7.66\% &2.30\% &0.77\% &8.81\% &3.45\% &1.15\% &15.71\% &\textbf{18.77\%} &1.15\% \\
\midrule
Overall &zero-shot &2.07\% &1.55\% &2.24\% &3.45\% &\textbf{3.62\%} &1.38\% &1.03\% &3.28\% &1.38\% \\
&refinement &9.66\% &3.97\% &8.28\% &\textbf{17.24\%} &16.03\% &5.69\% &11.55\% &16.72\% &3.45\% \\
\midrule
Improvement & &7.59\% &2.42\% &6.04\% &\textbf{13.79\%} &12.41\% &4.31\% &10.52\% &13.44\% &2.07\% \\
\bottomrule
\end{tabular}
\end{adjustbox}
\end{table}

Table~\ref{tab:relaxoverall} shows the results on \verirelax. Similarly, at the first attempt, absolute numbers remain low (less than 4\%) for all models. At the fifth iteration, o4-mini tops the competition with 17.24\%, followed closely by Deepseek (16.72\%), Claude (16.03\%), and Llama (11.55\%). Feedback leads to substantial improvement for most models, achieving relative gains of over 10\%.

In conclusion, while few models succeed from scratch, many become competitive when guided by partial context. Open-source models like Deepseek, and Llama outperform many closed-source counterparts, showing strong potential for real-world deployment in assisted formal verification. These results also underscore the importance of combining structural hints, feedback loops, and domain-specific strengths when applying LLMs to formal reasoning tasks.

\fbox{\parbox{\linewidth}{\textbf{Key Insights}: Average pass rates for all evaluated models remain low at 10\% on \verify and 18\% on \verirelax, revealing the challenges formal verification poses even to SOTA LLMs. All models improve with feedback.}}

\vspace{-.25cm}

\subsection{Failure Mode Distribution}
\vspace{-.3cm}
\begin{figure}[htbp]
  \centering
  \begin{minipage}[b]{0.49\textwidth}
    \centering
\includegraphics[width=\linewidth]{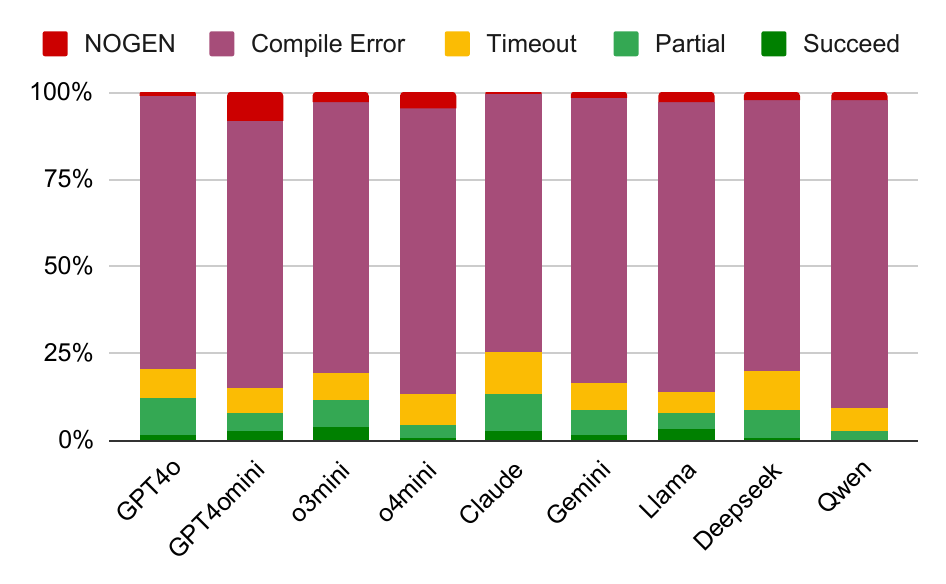}
    \vspace{-.8cm}
    \caption{zero-shot on VerifyThisBench}
    \label{fig:fig1}
  \end{minipage}
  \hfill
  \begin{minipage}[b]{0.49\textwidth}
    \centering
\includegraphics[width=\linewidth]{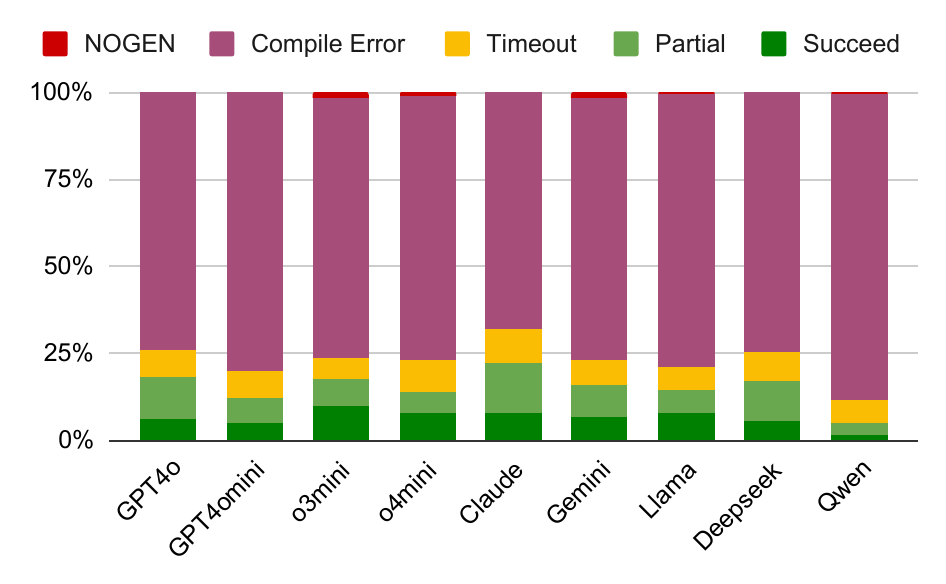}
    \vspace{-.8cm}
    \caption{refinement on VerifyThisBench}
    \label{fig:fig2}
  \end{minipage}
  \begin{minipage}[b]{0.49\textwidth}
    \centering
\includegraphics[width=\linewidth]{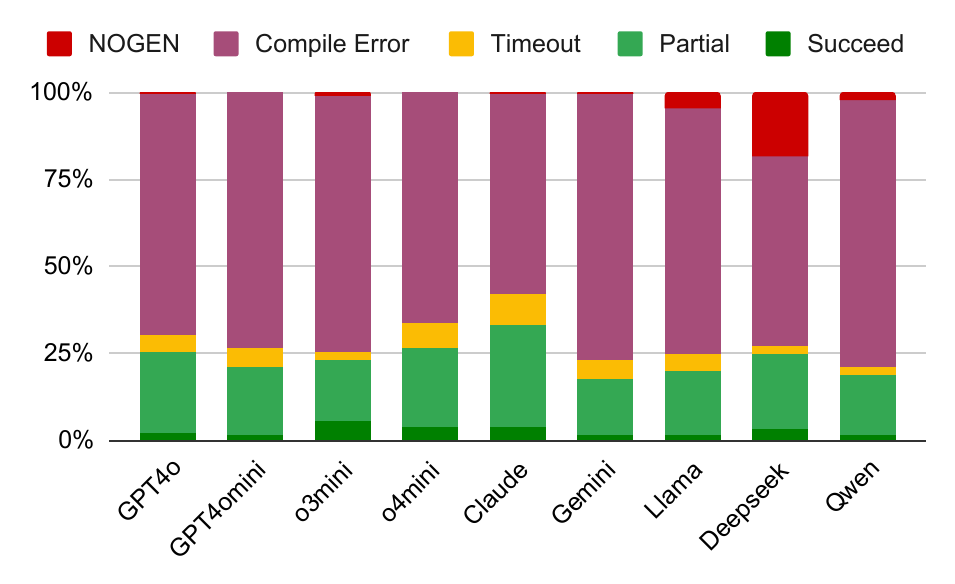}
    \vspace{-.8cm}
    \caption{zero-shot on VerifyThisBenchXS}
    \label{fig:fig3}
  \end{minipage}
  \hfill
  \begin{minipage}[b]{0.49\textwidth}
    \centering
\includegraphics[width=\linewidth]{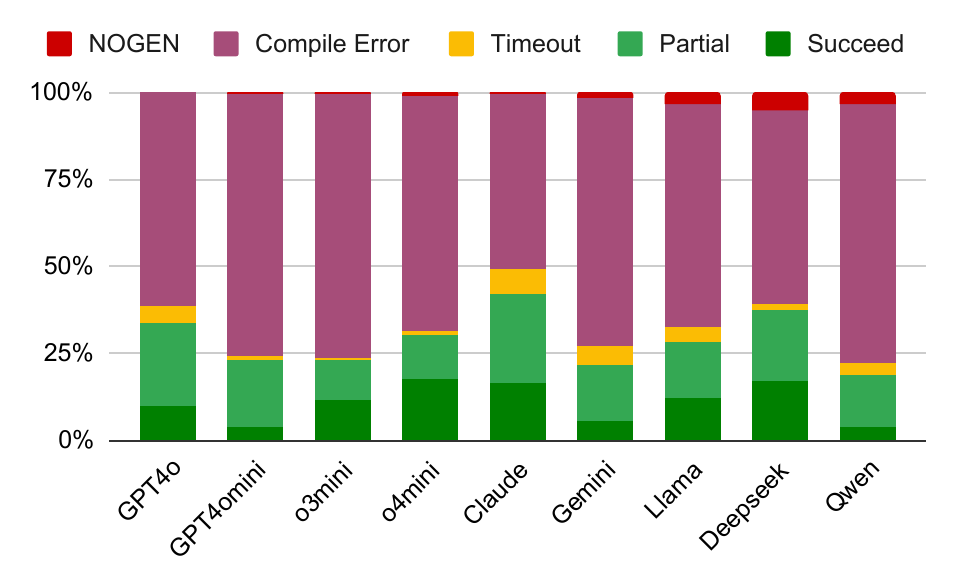}
    \vspace{-.8cm}
    \caption{refinement on VerifyThisBenchXS}
    \label{fig:fig4}
  \end{minipage}

\end{figure}

We categorize outcomes as \textbf{NOGEN} (no code detected), \textbf{Compile Error}, \textbf{Timeout} (compiles but exceeds verifier time budget), \textbf{Partial} (some but not all obligations proved), and \textbf{Succeed}.
Figures~\ref{fig:fig1} to \ref{fig:fig4} show clear improvements in model's performance when partial solution templates are provided in the relaxed settings. 
% Note that the left column represents zero-shot(i.e. Iter 1) and the right column shows refinement at Iter 5.

Specifically, \textbf{partial} success rates increase significantly, indicating that template hints help models generate more accurate solutions. \textbf{Timeout} rates remain relatively stable. This state indicates that models are making meaningful progress toward valid proofs, but the verifier struggles to find counterexamples on difficult obligations. \textbf{Compilation errors} still dominate but tend to decrease under the relaxed setting for some models, demonstrating that not needing to generate from scratch helps reduce syntax-level mistakes. However, some models like GPT4o-mini and o3-mini exhibit mixed trends, suggesting that while the template helps, the model’s internal understanding and code generation fidelity still vary.

If we relax the metric to consider compilable code rather than fully verified solutions, Claude, GPT-4o, and Deepseek consistently emerge as the top performers across both benchmarks. Notably, Claude generates compilable outputs in nearly 50\% of attempts on \verirelax and around 25\% on \verify in the first attempt alone, highlighting its strong baseline capability even without iterative feedback.
\vspace{-.2cm}
\fbox{\parbox{\linewidth}{\textbf{Key Insights}: While compilation error dominates in both benchmarks, in the relax setting we observe decreases in such failures and increases of partial correct or compilable solutions, moving model performance closer to usable verification outputs even when full correctness is not achieved.}}

\subsection{Coherence}

% add coherence by tool
%  remove first two rows

Table~\ref{tab:coh} reports each model’s coherence confidence, i.e. whether the model believes its generated specification matches the intended problem requirement. Importantly, this “self” alignment assessment is computed in a separate pass without chain‑of‑thought disclosure of how the answer was generated by the model and is thus a statistically independent evaluation. This metric is evaluated across the verified fraction of the outputs. While passing a formal verifier indicates syntactic and logical correctness, it does not address the alignment problem (i.e., whether the verified implementation perfectly aligns with the user-intent expressed in natural language descriptions); hence, coherence offers complementary insight. Notably, except o3-mini and Qwen, models's confidence is less than 50\% on passed solutions.

The results reveal considerable variance across models in their self-assessment behavior. Models like o3-mini and Claude exhibit high confidence, often reporting over 80\% coherence even in the zero-shot setting, suggesting strong internal certainty—though this may reflect overconfidence rather than accurate introspection. In contrast, models like GPT-4o and Llama show much more conservative estimates, with coherence below 30\%, indicating either better-calibrated uncertainty or limited self-awareness. Interestingly, refinement tends to reduce overconfidence for some models (e.g., Claude) while slightly improving coherence calibration for others (e.g., GPT-4o and Deepseek), suggesting iterative attempts help align perceived and actual correctness. 
\vspace{-.2cm}
\begin{table}[!htp]\centering
\caption{Self-Assessment of Specification Coherence on \verify}\label{tab:coh}
\scriptsize
\begin{tabular}{lrrrrrrrrrrr}\toprule
& Attempt &GPT4o &GPT4o-mini &o3-mini &o4-mini &Claude &Gemini &Llama &Deepseek &Qwen \\\midrule
CBMC &zero-shot &15.38\% & 0\% &84.62\% & 0\% &80.00\% &33.33\% & 0\% & 100\% & 100\% \\
&refinement &16.13\% & 0\% &61.54\% &12.50\% &26.47\% &13.64\% &3.23\% &20.59\% & 100\% \\\midrule
Dafny &zero-shot &- &- & 100\% & 100\% & 100\% &- &- & 100\% &- \\
&refinement &50.00\% & 100\% & 100\% &62.50\% &76.47\% &50.00\% &25.00\% & 100\% & 0\% \\\midrule
Frama-C &zero-shot &- & 100\% &- &- & 100\% & 0\% &- & 100\% &- \\
&refinement & 100\% & 100\% &75.00\% &80.00\% &70.59\% &53.85\% & 0\% & 100\% &- \\\midrule
VerCors &zero-shot &- & 100\% & 100\% &33.33\% &- &66.67\% &69.23\% & 100\% &- \\
&refinement &66.67\% & 100\% & 100\% &62.50\% & 0\% &46.15\% &61.11\% &85.71\% &80.00\% \\\midrule
VeriFast &zero-shot &- &- &- &- &- &- &- &- &- \\
&refinement &- &- &- & 0\% &- &- & 0\% & 100\% &- \\\midrule
Verus &zero-shot & 0\% &30.00\% & 100\% & 0\% & 0\% & 0\% &8.33\% & 0\% & 0\% \\
&refinement & 0\% &28.57\% &93.94\% &7.69\% & 0\% & 0\% &3.70\% & 0\% & 0\% \\\midrule
Why3 &zero-shot &- &- & 100\% & 100\% & 100\% & 0\% &- & 100\% &- \\
&refinement &- &- & 100\% &53.33\% &35.71\% &22.22\% & 0\% & 100\% &- \\\midrule
Average &zero-shot &12.50\% &25.00\% &94.87\% &50.00\% &88.00\% &43.75\% &27.78\% &90.91\% &66.67\% \\
&refinement &28.36\% &20.00\% &82.00\% &36.47\% &45.35\% &34.25\% &16.47\% &46.43\% &75.00\% \\
\bottomrule
\end{tabular}
\end{table}

We manually inspected a subset of successful solutions to validate if generated specifications align with the intended problem. Except for o3-mini, most models appear honest in their coherence self-assessments, with no false negatives found. Thus, our evaluation reflects an optimistic upper bound on true alignment—assuming coherence estimates are accurate and verifier passes indicate best-case correctness.
% \comment{unclear meaning to me}. \xun{two cases: if the passed solution's spec can align with user intent or not. the passing rate here is an upperbound of solutions that both (pass verifier + align user intent)}
\iclr{Automatically verifying the alignment between generated specifications and user intent in natural language remains an open technical challenge~\cite{lahiri2024evaluatingllmdrivenuserintentformalization}.  Our benchmark serves as a valuable resource for systematically investigating this specification–intent alignment problem in future research. In addition, we explore a test-based evaluation approach, with preliminary results presented in Appendix~\ref{app:testexp}} 
% \comment{if we want to link spec test agent Appn H to here}
% \xun{We explore testing-based evaluation and preliminary result is shown in Appendix H.}
% We manually inspected a subset of successful solutions and found few false negatives, suggesting that most models’ self‑assessments are not deceptively optimistic on this subset. Overall, coherence provides an optimistic upper bound on semantic alignment and highlights gaps not captured by pure verification outcomes.

\fbox{\parbox{\linewidth}{\textbf{Key Insights}: Models show a wide range in coherence confidence level, suggesting varied internal behaviors. On average, only 43\% of passed solutions are judged coherent and our manual review suggests strong alignment.}}

\subsection{Performance by Tools}\label{sec:tool}

Table~\ref{tab:tool} shows that all tools benefit from iterative refinement through feedback. In the \verify~setting, CBMC and Verus exhibit the most pronounced improvements, likely due to their syntactic resemblance to C and Rust, making them more accessible to language models. Dafny also shows moderate gains in this setting. In \verirelax, improvements are even more substantial. Dafny, in particular, demonstrates a leap from near-zero success rate to over 21.4\%; Verus observes an improvement around 10\%. In contrast, tools such as VeriFast, Frama-C, and Why3 remain largely stagnant on both benchmarks, suggesting either stricter syntactic or semantic constraints, or a structural mismatch with current model capabilities.
\vspace{-.2cm}
\begin{table}[!htp]\centering
\caption{Average Pass Rates across Tools}\label{tab:tool}
\scriptsize
\begin{tabular}{lrrrrrrrrr}\toprule
&Attempt &CBMC &Dafny &Frama-C &VerCors &VeriFast &Verus &Why3 \\\midrule
\multirow{2}{*}{\verify} &zero-shot &\textbf{4.76\%} &1.30\% &0.79\% &2.31\% &0 &3.32\% &0.51\% \\
&refinement &\textbf{18.11\%} &4.47\% &4.26\% &5.34\% &0.43\% &8.15\% &3.75\% \\\midrule
\multirow{2}{*}{\verirelax} &zero-shot &- &1.35\% &0.82\% &0 &5.22\% &\textbf{7.52}\% &0.21\% \\
&refinement &- &\textbf{21.47\%} &4.53\% &1.28\% &9.93\% &17.28\% &6.64\% \\
\bottomrule
\end{tabular}
\end{table}

\subsection{Performance by Relaxation}
\vspace{-.2cm}
\begin{table}[!htp]\centering
\caption{Overall Performance across Different Relaxation Settings in \verirelax}\label{tab:byrelax}
\scriptsize
\begin{tabular}{lrrrrrrr}\toprule
&\multicolumn{2}{c}{Code} &\multicolumn{2}{c}{Specification} &\multicolumn{2}{c}{Loop} \\\cmidrule{2-7}
Model &Zero-shot &Refinement &Zero-shot &Refinement &Zero-shot &Refinement \\\midrule
GPT4o &0.88\% &11.06\% &3.00\% &9.87\% &2.48\% &6.61\% \\
GPT4omini &0.88\% &3.98\% &2.15\% &3.86\% &1.65\% &4.13\% \\
o3mini &0.88\% &7.52\% &2.58\% &7.72\% &4.13\% &10.74\% \\
o4mini &0.88\% &14.16\% &\textbf{5.15\%} &18.45\% &4.96\% &\textbf{20.66\%} \\
Claude &\textbf{2.21\%} &15.04\% &4.29\% &\textbf{19.31\%} &4.96\% &11.57\% \\
Gemini &1.33\% &6.19\% &1.29\% &4.72\% &1.65\% &6.61\% \\
Llama &0.44\% &11.95\% &1.72\% &12.88\% &0.83\% &8.26\% \\
Deepseek &0.44\% &\textbf{15.49\%} &4.72\% &\textbf{19.31\%} &\textbf{5.79\%} &14.05\% \\
Qwen &1.33\% &3.54\% &1.29\% &3.86\% &1.65\% &2.48\% \\
\midrule
Overall &1.05\% &9.73\% &2.90\% &\textbf{11.27\%} &3.20\% &9.81\% \\

\bottomrule
\end{tabular}
\end{table}

Table~\ref{tab:byrelax} breaks down \verirelax results by \textbf{Code‑Gen}, \textbf{Spec‑Gen}, and \textbf{Loop‑Gen}. Iterative refinement consistently improves pass rates across all categories.
%Table~\ref{tab:byrelax} categorizes performance based on three relaxation types: 1) specification-gen, where the implementation and proof are given and the model fills in the spec; 2) code-gen, where the spec is provided and the model completes both the implementation and proof; and 3) loop-gen, where the full solution is available except for the loop invariant, which the model must supply. Across all categories, results indicate that iterative refinement or feedback significantly aids verification success.
%\comment{the definitions are given before, consider shorten if no space}

Among the three, spec-gen yields the highest overall pass rates, suggesting that models can more readily articulate reasoning about what a program is supposed to do, given a working implementation and its proof context. Completing loop invariant, arguably the most abstract and logically demanding task, results in pass rate lower than 10\%, though still showing solid gains with retries. This points to the inherent difficulty models face in understanding and completing partial proofs. 

% Completing code implementation falls between the two, showing that models can sometimes generate plausible code.

\fbox{\parbox{0.98\linewidth}{\textbf{Key Insights}: Generating the entire solution holistically (overall pass rate@9.73\%) may not be more difficult than generating a specific one, e.g., loop invariant (overall pass rate@9.81\%).}}

%Moreover, \textsc{VeriRelax} appears particularly helpful for tools like Frama-C and Why3, where specialized syntax and rigorous specification logic previously caused models to fail. Overall, the contrast underscores how partial automation and interactive programming environments can unlock latent capabilities in current LLMs, even when their zero-shot abilities remain limited.

\section{Conclusion}
In this work, we introduce \textbf{\verify} and \textbf{\verirelax} to evaluate the formal verification capabilities of large language models, systematically assessing their performance across a range of tools, tasks, and relaxation settings. Despite the use of SOTA models, results show generally poor performance, particularly in strict end-to-end settings that require complete formal reasoning without assistance. These findings highlight significant gaps in current models’ ability to generate semantically and logically correct solutions in formal domains.

% Backup version:
% VerifyThisBench evaluates LLMs on end‑to‑end program verification from natural language, requiring specification synthesis, implementation, and proofs that compile and verify. Despite iterative feedback, current models achieve single‑digit pass rates in the strict setting and teens in the relaxed setting, revealing significant gaps in formal reasoning. Limitations include reliance on model self‑assessment for coherence and exploration of only feedback‑based refinement. Future work should investigate verifier‑in‑the‑loop decoding, dedicated fine‑tuning on formally verified corpora, invariant‑guided prompting, and cross‑tool generalization to advance the state of verified program generation.

%remove this 
% This study also has several limitations. Our coherence metric relies on the model's self-assessment, which may not reliably reflect true specification alignment. Additionally, our exploration of iteration is limited to feedback-based refinement; more sophisticated approaches, such as fine-tuning on formal verification tasks or incorporating external verification guidance, may yield further insights and improvements. Future work should explore these directions to better understand and advance the formal reasoning capabilities of language models.
%\section*{Acknowledgments}
%This was was supported in part by......
\bibliographystyle{plain}  
\bibliography{main}  

\begin{thebibliography}{10}

\bibitem{claude2024introducingV3}
Anthropic.
\newblock Introducing the next generation of claude, 2024.

\bibitem{claude37}
Anthropic.
\newblock Claude 3.7 sonnet and claude code, 2025.

\bibitem{austin2021programsynthesislargelanguage}
Jacob Austin, Augustus Odena, Maxwell Nye, Maarten Bosma, Henryk Michalewski, David Dohan, Ellen Jiang, Carrie Cai, Michael Terry, Quoc Le, and Charles Sutton.
\newblock Program synthesis with large language models, 2021.

\bibitem{barbosa2022cvc5}
Haniel Barbosa, Clark Barrett, Martin Brain, Gereon Kremer, Hanna Lachnitt, Makai Mann, Abdalrhman Mohamed, Mudathir Mohamed, Aina Niemetz, Andres N{\"o}tzli, Alex Ozdemir, Mathias Preiner, Andrew Reynolds, Ying Sheng, Cesare Tinelli, and Yoni Zohar.
\newblock cvc5: A versatile and industrial-strength smt solver.
\newblock In Dana Fisman and Grigore Rosu, editors, {\em Tools and Algorithms for the Construction and Analysis of Systems}, pages 415--442, Cham, 2022. Springer International Publishing.

\bibitem{rankInv23}
Saikat Chakraborty, Shuvendu~K. Lahiri, Sarah Fakhoury, Akash Lal, Madanlal Musuvathi, Aseem Rastogi, Aditya Senthilnathan, Rahul Sharma, and Nikhil Swamy.
\newblock Ranking llm-generated loop invariants for program verification.
\newblock In Houda Bouamor, Juan Pino, and Kalika Bali, editors, {\em Findings of the Association for Computational Linguistics: {EMNLP} 2023, Singapore, December 6-10, 2023}, pages 9164--9175. Association for Computational Linguistics, 2023.

\bibitem{chen2021evaluatinglargelanguagemodels}
Mark Chen, Jerry Tworek, Heewoo Jun, Qiming Yuan, Henrique~Ponde de~Oliveira~Pinto, Jared Kaplan, Harri Edwards, Yuri Burda, Nicholas Joseph, Greg Brockman, Alex Ray, Raul Puri, Gretchen Krueger, Michael Petrov, Heidy Khlaaf, Girish Sastry, Pamela Mishkin, Brooke Chan, Scott Gray, Nick Ryder, Mikhail Pavlov, Alethea Power, Lukasz Kaiser, Mohammad Bavarian, Clemens Winter, Philippe Tillet, Felipe~Petroski Such, Dave Cummings, Matthias Plappert, Fotios Chantzis, Elizabeth Barnes, Ariel Herbert-Voss, William~Hebgen Guss, Alex Nichol, Alex Paino, Nikolas Tezak, Jie Tang, Igor Babuschkin, Suchir Balaji, Shantanu Jain, William Saunders, Christopher Hesse, Andrew~N. Carr, Jan Leike, Josh Achiam, Vedant Misra, Evan Morikawa, Alec Radford, Matthew Knight, Miles Brundage, Mira Murati, Katie Mayer, Peter Welinder, Bob McGrew, Dario Amodei, Sam McCandlish, Ilya Sutskever, and Wojciech Zaremba.
\newblock Evaluating large language models trained on code, 2021.

\bibitem{demoura2008z3}
Leonardo de~Moura and Nikolaj Bj{\o}rner.
\newblock Z3: An efficient smt solver.
\newblock In {\em Proceedings of the 14th International Conference on Tools and Algorithms for the Construction and Analysis of Systems (TACAS 2008)}, pages 337--340. Springer, 2008.

\bibitem{deMoura2018lean}
Leonardo de~Moura, Soonho Kong, Jeremy Avigad, Floris~Van Doorn, and Jakob von Raumer.
\newblock {The Lean Theorem Prover (system description)}.
\newblock 6 2018.

\bibitem{gemini}
Google DeepMind.
\newblock Gemini-2.5-flash, 2025.

\bibitem{deepseekai2024deepseekv3technicalreport}
DeepSeek-AI.
\newblock Deepseek-v3 technical report, 2024.

\bibitem{10.1007/978-3-031-67695-6_5}
Xavier Denis and Stephen~F. Siegel.
\newblock Verifythis 2023: An international program verification competition.
\newblock In {\em TOOLympics Challenge 2023: Updates, Results, Successes of the Formal-Methods Competitions}, page 147–159, Berlin, Heidelberg, 2024. Springer-Verlag.

\bibitem{dijkstra1972humble}
Edsger~W. Dijkstra.
\newblock The humble programmer.
\newblock Technical Report EWD340, EWD, 1972.
\newblock Technical report from the EWD series.

\bibitem{LLMpostcondition24}
Madeline Endres, Sarah Fakhoury, Saikat Chakraborty, and Shuvendu~K. Lahiri.
\newblock Can large language models transform natural language intent into formal method postconditions?
\newblock {\em Proc. {ACM} Softw. Eng.}, 1({FSE}):1889--1912, 2024.

\bibitem{AILuminate25}
Shaona Ghosh, Heather Frase, Adina Williams, Sarah Luger, Paul R{\"{o}}ttger, Fazl Barez, Sean McGregor, Kenneth Fricklas, Mala Kumar, Quentin Feuillade{-}Montixi, Kurt Bollacker, Felix Friedrich, Ryan Tsang, Bertie Vidgen, Alicia Parrish, Chris Knotz, Eleonora Presani, Jonathan Bennion, Marisa~Ferrara Boston, Mike Kuniavsky, Wiebke Hutiri, James Ezick, Malek~Ben Salem, Rajat Sahay, Sujata~S. Goswami, Usman Gohar, Ben Huang, Supheakmungkol Sarin, Elie Alhajjar, Canyu Chen, Roman Eng, Kashyap~Ramanandula Manjusha, Virendra Mehta, Eileen Long, Murali Emani, Natan Vidra, Benjamin Rukundo, Abolfazl Shahbazi, Kongtao Chen, Rajat Ghosh, Vithursan Thangarasa, Pierre Peign{\'{e}}, Abhinav Singh, Max Bartolo, Satyapriya Krishna, Mubashara Akhtar, Rafael Gold, Cody Coleman, Luis Oala, Vassil Tashev, Joseph~Marvin Imperial, Amy Russ, Sasidhar Kunapuli, Nicolas Miailhe, Julien Delaunay, Bhaktipriya Radharapu, Rajat Shinde, Tuesday, Debojyoti Dutta, Declan Grabb, Ananya Gangavarapu, Saurav Sahay, Agasthya Gangavarapu,
  Patrick Schramowski, Stephen Singam, Tom David, Xudong Han, Priyanka~Mary Mammen, Tarunima Prabhakar, Venelin Kovatchev, Ahmed Ahmed, Kelvin~N. Manyeki, Sandeep Madireddy, Foutse Khomh, Fedor Zhdanov, Joachim Baumann, Nina Vasan, Xianjun Yang, Carlos Mougn, Jibin~Rajan Varghese, Hussain Chinoy, Seshakrishna Jitendar, Manil Maskey, Claire~V. Hardgrove, Tianhao Li, Aakash Gupta, Emil Joswin, Yifan Mai, Shachi~H. Kumar, Cigdem Patlak, Kevin Lu, Vincent Alessi, Sree~Bhargavi Balija, Chenhe Gu, Robert Sullivan, James Gealy, Matt Lavrisa, James Goel, Peter Mattson, Percy Liang, and Joaquin Vanschoren.
\newblock {AILuminate}: Introducing v1.0 of the {AI} risk and reliability benchmark from mlcommons.
\newblock {\em CoRR}, abs/2503.05731, 2025.

\bibitem{copilot2021introducing}
{GitHub}.
\newblock {Introducing GitHub Copilot}.
\newblock \url{https://github.blog/2021/06/29/introducing-github-copilot}, 2021.
\newblock Blog post announcing GitHub Copilot.

\bibitem{google2024gemini}
Google.
\newblock Gemini 1.5: Unlocking multimodal understanding across millions of tokens of context, 2024.

\bibitem{CRUXEval24}
Alex Gu, Baptiste Rozi{\`{e}}re, Hugh~James Leather, Armando Solar{-}Lezama, Gabriel Synnaeve, and Sida Wang.
\newblock {CRUXEval}: {A} benchmark for code reasoning, understanding and execution.
\newblock In {\em Forty-first International Conference on Machine Learning, {ICML} 2024, Vienna, Austria, July 21-27, 2024}. OpenReview.net, 2024.

\bibitem{hendrycks2021measuringcodingchallengecompetence}
Dan Hendrycks, Steven Basart, Saurav Kadavath, Mantas Mazeika, Akul Arora, Ethan Guo, Collin Burns, Samir Puranik, Horace He, Dawn Song, and Jacob Steinhardt.
\newblock Measuring coding challenge competence with apps, 2021.

\bibitem{huth2004logic}
Michael Huth and Mark Ryan.
\newblock {\em Logic in Computer Science: Modelling and Reasoning about Systems}.
\newblock Cambridge University Press, 2004.

\bibitem{jimenez2024swebenchlanguagemodelsresolve}
Carlos~E. Jimenez, John Yang, Alexander Wettig, Shunyu Yao, Kexin Pei, Ofir Press, and Karthik Narasimhan.
\newblock Swe-bench: Can language models resolve real-world github issues?, 2024.

\bibitem{kamath2023finding}
Adharsh Kamath, Aditya Senthilnathan, Saikat Chakraborty, Pantazis Deligiannis, Shuvendu~K Lahiri, Akash Lal, Aseem Rastogi, Subhajit Roy, and Rahul Sharma.
\newblock Finding inductive loop invariants using large language models.
\newblock {\em arXiv preprint arXiv:2311.07948}, 2023.

\bibitem{DynaBench21}
Douwe Kiela, Max Bartolo, Yixin Nie, Divyansh Kaushik, Atticus Geiger, Zhengxuan Wu, Bertie Vidgen, Grusha Prasad, Amanpreet Singh, Pratik Ringshia, Zhiyi Ma, Tristan Thrush, Sebastian Riedel, Zeerak Waseem, Pontus Stenetorp, Robin Jia, Mohit Bansal, Christopher Potts, and Adina Williams.
\newblock Dynabench: Rethinking benchmarking in {NLP}.
\newblock In Kristina Toutanova, Anna Rumshisky, Luke Zettlemoyer, Dilek Hakkani{-}T{\"{u}}r, Iz~Beltagy, Steven Bethard, Ryan Cotterell, Tanmoy Chakraborty, and Yichao Zhou, editors, {\em Proceedings of the 2021 Conference of the North American Chapter of the Association for Computational Linguistics: Human Language Technologies, {NAACL-HLT} 2021, Online, June 6-11, 2021}, pages 4110--4124. Association for Computational Linguistics, 2021.

\bibitem{kirchner2015frama}
Florent Kirchner, Nikolai Kosmatov, Virgile Prevosto, Julien Signoles, and Boris Yakobowski.
\newblock Frama-c: A software analysis perspective.
\newblock volume~27, pages 573--609. Springer, 2015.

\bibitem{kroening2023cbmccboundedmodel}
Daniel Kroening, Peter Schrammel, and Michael Tautschnig.
\newblock {CBMC}: The {C} bounded model checker, 2023.

\bibitem{lahiri2024evaluatingllmdrivenuserintentformalization}
Shuvendu~K. Lahiri.
\newblock Evaluating llm-driven user-intent formalization for verification-aware languages, 2024.

\bibitem{lai2022ds1000naturalreliablebenchmark}
Yuhang Lai, Chengxi Li, Yiming Wang, Tianyi Zhang, Ruiqi Zhong, Luke Zettlemoyer, Scott~Wen tau Yih, Daniel Fried, Sida Wang, and Tao Yu.
\newblock Ds-1000: A natural and reliable benchmark for data science code generation, 2022.

\bibitem{lattuada2023verus}
Andrea Lattuada, Travis Hance, Chanhee Cho, Matthias Brun, Isitha Subasinghe, Yi~Zhou, Jon Howell, Bryan Parno, and Chris Hawblitzel.
\newblock Verus: Verifying rust programs using linear ghost types.
\newblock {\em Proceedings of the ACM on Programming Languages}, 7(OOPSLA1):286--315, 2023.

\bibitem{leino2010dafny}
K.~R.~M. Leino.
\newblock Dafny: An automatic program verifier for functional correctness.
\newblock In {\em Proceedings of the 2010 International Conference on Verification, Model Checking, and Abstract Interpretation (VMCAI)}, pages 348--352. Springer, 2010.

\bibitem{Li_2022}
Yujia Li, David Choi, Junyoung Chung, Nate Kushman, Julian Schrittwieser, Rémi Leblond, Tom Eccles, James Keeling, Felix Gimeno, Agustin Dal~Lago, Thomas Hubert, Peter Choy, Cyprien de~Masson~d’Autume, Igor Babuschkin, Xinyun Chen, Po-Sen Huang, Johannes Welbl, Sven Gowal, Alexey Cherepanov, James Molloy, Daniel~J. Mankowitz, Esme Sutherland~Robson, Pushmeet Kohli, Nando de~Freitas, Koray Kavukcuoglu, and Oriol Vinyals.
\newblock Competition-level code generation with alphacode.
\newblock {\em Science}, 378(6624):1092–1097, December 2022.

\bibitem{liu2023codegeneratedchatgptreally}
Jiawei Liu, Chunqiu~Steven Xia, Yuyao Wang, and Lingming Zhang.
\newblock Is your code generated by chatgpt really correct? rigorous evaluation of large language models for code generation, 2023.

\bibitem{loughridge2024dafnyBench}
Chloe Loughridge, Qinyi Sun, Seth Ahrenbach, Federico Cassano, Chuyue Sun, Ying Sheng, Anish Mudide, Md~Rakib~Hossain Misu, Nada Amin, and Max Tegmark.
\newblock Dafnybench: A benchmark for formal software verification.
\newblock {\em arXiv preprint arXiv:2406.08467}, 2024.

\bibitem{metallama}
Meta.
\newblock Llama3.3-70b-instruct.

\bibitem{nipkow2002isabelle}
Tobias Nipkow, Lawrence~C. Paulson, and Markus Wenzel.
\newblock {\em Isabelle/HOL: A Proof Assistant for Higher-Order Logic}, volume 2283 of {\em Lecture Notes in Computer Science}.
\newblock Springer, 2002.

\bibitem{openai2024gpt4o}
{OpenAI}.
\newblock {Hello GPT-4o}, 2024.

\bibitem{OpenAI}
OpenAI.
\newblock Openai model list, 2025.

\bibitem{LLMinv23}
Kexin Pei, David Bieber, Kensen Shi, Charles Sutton, and Pengcheng Yin.
\newblock Can large language models reason about program invariants?
\newblock In Andreas Krause, Emma Brunskill, Kyunghyun Cho, Barbara Engelhardt, Sivan Sabato, and Jonathan Scarlett, editors, {\em International Conference on Machine Learning, {ICML} 2023, 23-29 July 2023, Honolulu, Hawaii, {USA}}, volume 202 of {\em Proceedings of Machine Learning Research}, pages 27496--27520. {PMLR}, 2023.

\bibitem{qwen2.5}
Qwen.
\newblock Qwen2.5: A party of foundation models, September 2024.

\bibitem{si2020code2inv}
Xujie Si, Aaditya Naik, Hanjun Dai, Mayur Naik, and Le~Song.
\newblock Code2inv: A deep learning framework for program verification.
\newblock In {\em Computer Aided Verification: 32nd International Conference, CAV 2020, Los Angeles, CA, USA, July 21--24, 2020, Proceedings, Part II 32}, pages 151--164. Springer, 2020.

\bibitem{SVCOMP}
SV-COMP-org.
\newblock \url{https://sv-comp.sosy-lab.org/}.

\bibitem{sygus}
Sygus-org.
\newblock Sygus.
\newblock \url{https://sygus.org/}.

\bibitem{thakur2025clevercuratedbenchmarkformally}
Amitayush Thakur, Jasper Lee, George Tsoukalas, Meghana Sistla, Matthew Zhao, Stefan Zetzsche, Greg Durrett, Yisong Yue, and Swarat Chaudhuri.
\newblock Clever: A curated benchmark for formally verified code generation, 2025.

\bibitem{VercorsGithub}
VerCors.
\newblock Vercors tool.

\bibitem{VerifastGithub}
VeriFast.
\newblock Verifast tool.

\bibitem{verifythis}
{VerifyThis Competition Series}, 2025.

\bibitem{why3}
Why3.
\newblock Why3 project.

\bibitem{LLMinv-ASE24}
Guangyuan Wu, Weining Cao, Yuan Yao, Hengfeng Wei, Taolue Chen, and Xiaoxing Ma.
\newblock {LLM} meets bounded model checking: Neuro-symbolic loop invariant inference.
\newblock In Vladimir Filkov, Baishakhi Ray, and Minghui Zhou, editors, {\em Proceedings of the 39th {IEEE/ACM} International Conference on Automated Software Engineering, {ASE} 2024, Sacramento, CA, USA, October 27 - November 1, 2024}, pages 406--417. {ACM}, 2024.

\bibitem{xia2024top}
Chunqiu~Steven Xia, Yinlin Deng, and Lingming Zhang.
\newblock Top leaderboard ranking= top coding proficiency, always? evoeval: Evolving coding benchmarks via llm.
\newblock {\em arXiv preprint arXiv:2403.19114}, 2024.

\bibitem{yang2024verusBench}
Chenyuan Yang, Xuheng Li, Md~Rakib~Hossain Misu, Jianan Yao, Weidong Cui, Yeyun Gong, Chris Hawblitzel, Shuvendu Lahiri, Jacob~R Lorch, Shuai Lu, et~al.
\newblock Autoverus: Automated proof generation for rust code.
\newblock {\em arXiv preprint arXiv:2409.13082}, 2024.

\bibitem{ye2025verina}
Zhe Ye, Zhengxu Yan, Jingxuan He, Timothe Kasriel, Kaiyu Yang, and Dawn Song.
\newblock Verina: Benchmarking verifiable code generation.
\newblock {\em arXiv preprint arXiv:2505.23135}, 2025.

\end{thebibliography}

\newpage
\appendix
\newpage
\section{Composition of VerifyThisBenchXS}\label{relaxbytool}
Table~\ref{tab:RelaxStats} presents the composition of \verirelax, summarizing the number of verification tasks by tool and category. It includes counts of implementations, specifications, and loop-related completion tasks for six verification tools: Dafny, Frama-C, VerCors, Verifast, Why3, and Verus. In total, the benchmark comprises 580 tasks, with 226 implementations, 233 specifications, and 121 loop invariants related examples.
\begin{table}[!htp]\centering
\caption{Composition of \verirelax}\label{tab:RelaxStats}
\begin{tabular}{l|r|r|r|r}\toprule
Tool &Implementaion &Specification &Loop &Total \\\midrule
Dafny &28 &25 &21 &74 \\
Frama-C &15 &15 &24 &54 \\
VerCors &8 &8 &10 &26 \\
VeriFast &26 &31 &9 &66 \\
Why3 &118 &117 &26 &261 \\
Verus &31 & 27 & 31 & 99 \\ \midrule
Total &226 &233 &121 &580 \\
\bottomrule
\end{tabular}
\end{table}
% \section{Example system prompt}
% \input{appendix/examplesys}

% \section{Extra Data and Results}

\section{Model Versions}\label{version}
GPT-4o was evaluated using the version from August 6, 2024, while GPT-4o-mini and o4-mini correspond to the July 18, 2024 versions. The o3-mini model was accessed as of January 31, 2025. Claude refers to the Claude-3.7-Sonnet version released on February 24, 2025, and Gemini-2.5 Flash on the April 17, 2025 release. For open-source models, we used LLaMA3.3-70b Instruct from December 6, 2024, DeepSeek-chat-v3 from March 24, 2025, and Qwen2.5-72b Instruct from September 19, 2024. These version references ensure the reproducibility and consistency of our benchmarking results.

\section{Tool Versions}
\label{toolversion}
We report exact toolchain versions for reproducibility and summarize each tool’s verification model.
The Verus verifier was run using version v0.2025.04.03.0f22710, while Why3 was evaluated with version v1.6.0. For Frama-C, we used version v30.0, and VeriFast experiments were conducted with version v25.02. The Dafny toolchain ran on version v4.10.0, and VerCors with v2.3.0. Finally, we used CBMC version v6.5.0.

Docker container images and unified toolchain launch scripts are included in the released dataset.
Below we briefly describe each tool:

\begin{itemize}
  \item \textbf{Dafny:} A verification-aware programming language with built-in specification support (pre/post-conditions, invariants) and an automatic static verifier.
  \item \textbf{Why3:} A platform for deductive verification with its own intermediate language (WhyML) and integration with external theorem provers.
  \item \textbf{VeriFast:} A verifier for C and Java using separation logic, enabling modular reasoning about memory safety and functional correctness.
  \item \textbf{VerCors:} A verifier for concurrent programs in Java, C, and OpenCL, supporting permission-based separation logic and parallel reasoning.
  \item \textbf{Frama-C:} A modular analysis platform for C, using the ACSL specification language and combining abstract interpretation with deductive verification.
  \item \textbf{Verus:} A verifier for Rust programs that checks user-defined specifications using SMT solving, supporting low-level features and ownership semantics.
  \item \textbf{CBMC:} A bounded model checker for C and C++ that verifies safety and functional correctness by translating code into SAT/SMT formulas.
\end{itemize}

\section{Prompt Formats}\label{app:prompt}
As prompt optimization was not the focus of this work, we used a simple, uniform structure for all models to ensure fairness across different tools.
Each prompt consists of a system prompt describing the verification tool, followed by the problem description and task. System prompts used in our experiments are included in the released dataset (see artifact).

\textbf{(1) System prompt}: a concise tool description and key syntax/semantics reminders.
\begin{lstlisting} [frame=None]
You are an assistant that writes formally verified programs in <TOOL>.
- Use <language/syntax> with pre/postconditions, assertions, and loop invariants as required.
- The solution must compile and pass the <TOOL> verifier with a 60s timeout.
- Do not use unsupported features: <list>.
- Return a single <file-type>, with all annotations needed for verification.
\end{lstlisting}

\textbf{(2) User prompt}: the natural-language problem overview and the specific task.

\begin{lstlisting} [frame=None]
# Description
<Problem overview in natural language; may include pseudo-code.>

# Task
<Explicit instruction: implement/specify/prove/refine the desired property.>
\end{lstlisting}

\section{Statistics of \verify and \verirelax }\label{app:loc}

\begin{figure}[htp!]
    \centering
    \includegraphics[width=.8\linewidth]{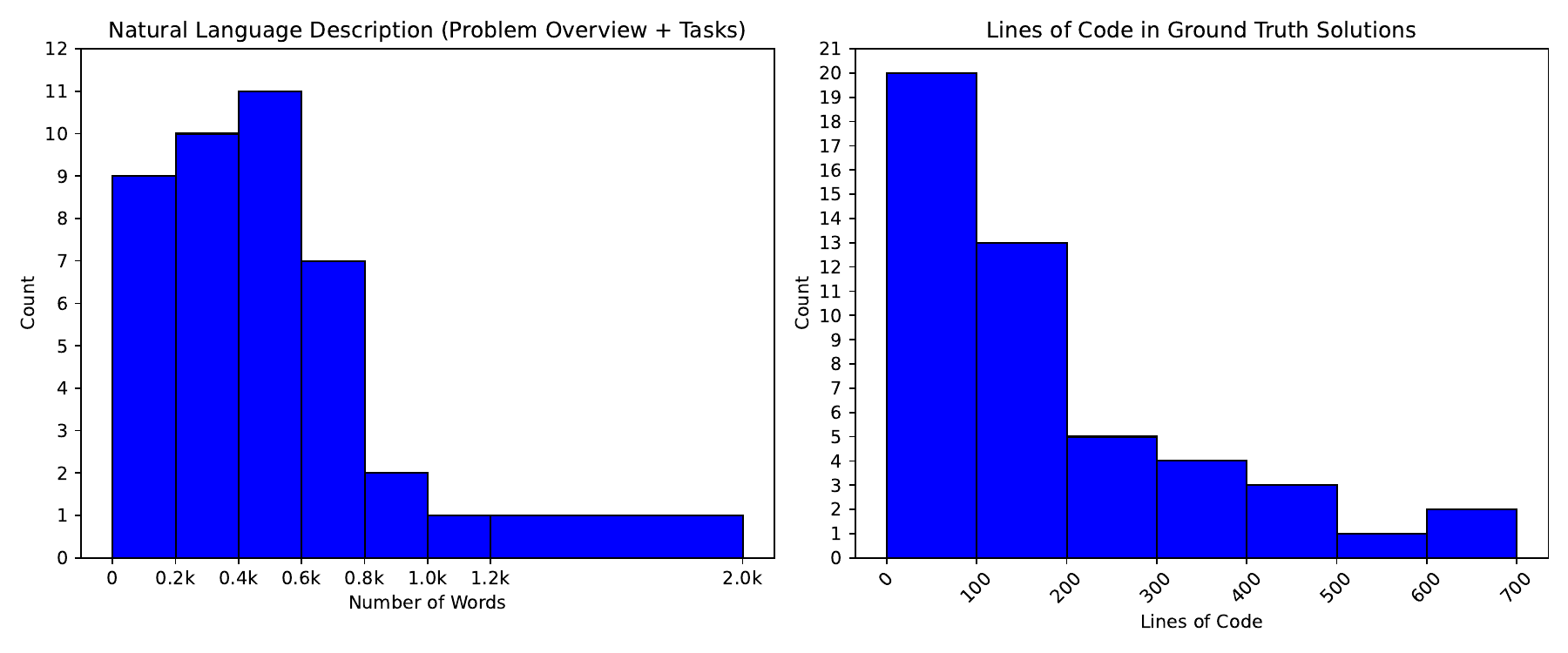}
    \caption{Distributions of dataset characteristics. (Left) Word count distribution of natural language descriptions for challenges. (Right) Line-of-code distribution of collected ground-truth solutions.}
    \label{fig:loc}
\end{figure}

To provide empirical support for our claim regarding the range of difficulty in the dataset, we report several descriptive statistics. The natural language descriptions (problem overview and task statements) vary substantially in length, with an average of 467 words, ranging from 30 to 1802 words. The distribution, shown in Figure~\ref{fig:loc} (left), indicates that most challenges fall within the 200–799 word range, with a small number extending beyond 1000 words. In terms of solution complexity, we analyzed 48 collected ground-truth implementations, which range from as short as 28 lines to 648 lines, with an average of 189.44 lines and a median of 124 lines per solution. As illustrated in Figure~\ref{fig:loc}  (right), the majority of these solutions are under 300 lines, with a few extending beyond 500 lines.

Beyond length, the diversity of task types also reflects difficulty variation: 15 out of 41 challenges involve relatively simple data structures such as binary trees and one-dimensional arrays, whereas the remaining challenges address more complex structures, including linked lists, graphs, queues, and specialized task-specific data types. Additionally, 11 out of 41 challenges explicitly require memory safety proofs, further illustrating the technical depth of the dataset.

Natural language descriptions in Verina~\cite{ye2025verina} have a median length of 110 and max length of 296 words, with accompanying code and specifications of up to 100 lines. Clever~\cite{thakur2025clevercuratedbenchmarkformally} reports proof lengths ranging from 10 to 225 lines. In contrast, our benchmark spans a much broader range of difficulty.

\section{Exploration of test-based specification verification}\label{app:testexp}
Inspired by parallel work~\cite{ye2025verina}, we further explore a test-based proxy to evaluate specification alignment. 
We manually construct desired input-output pairs of a problem, and verify them against the specifications generated by the models. We check if the \textit{inputs} imply the described \textit{pre-conditions}, and if the \underline{outputs} satisfy the \underline{post-conditions}.
Our setup supports open-ended, complex verification problems, without restrictions on how the function signatures or data structures are defined. As a preliminary experiment, we evaluated on all passed or failed samples of Dafny specifications generated from the \verify end-to-end tasks using test cases, on the following two benchmark problems:\\
\hspace*{10pt}\quad 1.\quad Finding the maximum in an array, and \\
\hspace*{10pt}\quad 2.\quad Finding the maximum in a tree. 

For the array version, 87.5\% of the generated specifications passed the test cases (as a reference, the model’s self-assessment on coherence: 93\%). For the tree version, only 10\% passed, mainly due to syntax errors, helper function verifiability, and other issues (reference on the model’s self-assessment on coherence: 87\%).

These results differ from our manual evaluations and the model’s self-assessments. Model’s assessment focuses on intent alignment, whereas testing requires functional correctness. This illustrates the complementary nature of different evaluation methods. 

\section{VerifyThisBenchXS Data Source}
Table~\ref{tab:source} lists the sources of solutions used to construct \verirelax. It includes the year of publication, the name of the verification challenge, the verification tool used, and the authors or contributors of each solution. We include canonical community solutions where available; in addition to the list, we contribute new Verus solutions (see Section 3.4).

\begin{table}[!htp]\centering
\caption{Solution used to generate \verirelax.}\label{tab:source}
\scriptsize
\begin{tabular}{lrrrr}\toprule
Year &Challenge Name & Tool &Authors \\\midrule
2024 &The Rope Data Structure &Why3 &Jean-Christophe Filliâtre \\
2024 &Smart Array Copy by Shuffled Subsegments &Why3 &Jean-Christophe Filliâtre \\
2023 &Binary Decision Diagrams &Why3 &Martin Clochard and Yannick Moy \\
2021 &Lexicographic Permutations &Why3&Jean-Christophe Filliâtre and Andrei Paskevich \\
2021 &Lexicographic Permutations &VerCors	& Marieke Huisman and Sebastiaan Joosten\\
2021 &DLL to BST &Why3 &Jean-Christophe Filliâtre and Andrei Paskevich \\
2021 &Shearsort &Why3 &Jean-Christophe Filliâtre and Andrei Paskevich \\
2019 &Monotonic Segments and GHC sort &Frama-C &Virgile Prevosto and Virgile Robles \\

2019 &Monotonic Segments and GHC sort &Dafny & Sample answer from report\\
2019 &Cartesian Trees &Frama-C &Virgile Prevosto and Virgile Robles \\
2019 &Sparse Matrix Multiplication &Frama-C &Virgile Prevosto and Virgile Robles \\
2018 &Array Based Queuing Lock &Why3 &Raphael Rieu \\
2018 &Gap buffer &Why3 &Raphael Rieu \\
2018 &Colored tiles &Why3 &Raphael Rieu \\
2017 &Pair Insertion Sort &Frama-C&Lionel Blatter and Jean-Christophe Léchenet \\
2017 &Pair Insertion Sort & Dafny &	Jon Mediero Iturrioz\\
2017 &Pair Insertion Sort & VerCors	& Marieke Huisman, Wytse Oortwijn \\
2017 &Maximum-sum Array(one-dimension) &Frama-C &Lionel Blatter and Jean-Christophe Léchenet \\
2017 &Odd-even Transposition Sort &Frama-C &Lionel Blatter and Jean-Christophe Léchenet \\
2017 &Tree Buffer &Frama-C &Lionel Blatter and Jean-Christophe Léchenet \\
2017 &Tree Buffer & VerCors	 & Marieke Huisman, Wytse Oortwijn\\
2016 &Matrix Multiplication &VeriFast &Bart Jacobs \\
2016 &Matrix Multiplication &Dafny	&Luca Weibel and Christiaan Dirkx \\
2016 &Matrix Multiplication &Why3&Martin Clochard and Léon Gondelman and Mário Pereira\\

2016 &Binary Tree Traversal &VeriFast &Bart Jacobs \\
2016 &Binary Tree Traversal &Why3 	&Martin Clochard and Léon Gondelman and Mário Pereira\\
2016 &Static Tree Barrier &VeriFast &Bart Jacobs \\
2015 &RELAXED PREFIX &Why3&Jean-Christophe Filliâtre and Guillaume Melquiond \\
2015 &PARALLEL GCD &Why3 &Jean-Christophe Filliâtre and Guillaume Melquiond \\
2015 &DANCING LINKS &Why3 &Jean-Christophe Filliâtre and Guillaume Melquiond \\
2012 &Longest Common Prefix &VeriFast &Bart Jacobs and Jan Smans \\
2012 &Prefix-Sum &VeriFast &Bart Jacobs and Jan Smans \\
2012 &Tree Del &VeriFast &Bart Jacobs and Jan Smans \\
2011 &Finding the Maximum in an Array &Dafny &Julian Tschannen and Nadia Polikarpova \\
2011 &Finding the Maximum in a Tree &Dafny &Julian Tschannen and Nadia Polikarpova \\
2011 &Finding Two Duplets in an Array &Dafny &Julian Tschannen and Nadia Polikarpova \\
\bottomrule
\end{tabular}
\end{table}

\newpage
\section{Example Challenge and Solution}\label{app:example}

\begin{figure}[h!]
\begin{lstlisting}[frame=none]
// # Description
// This challenge is an instance of Kaldewaij's Search by Elimination, where an element with a given property is located by eliminating elements that do not have that property. The challenge was selected as it involves a relatively simple but interesting invariant, expressing that the maximal element is in the remaining search space rather than maintaining the maximal element found so far. A pseudo-code implementation is as follows:
// int max(int[] a) {
//     int x = 0;
//     int y = a.length-1;
//     while (x != y) {
//       if (a[x] <= a[y]) x++;
//       else y--;
//     }
//     return x;
//  }
// # Task
// Please implement and verify that the index returned by the method max() given points to an element maximal in the array
pub fn max(a: &[int]) -> (result: usize)
\end{lstlisting}
\vspace{-\baselineskip}
\begin{lstlisting}[frame=none, firstnumber=15, backgroundcolor=\color{codeyellow}]
    requires
        a.len() > 0,
    ensures ({
        result < a.len(),
        forall|i: int| 0 <= i && i < a.len() ==> a[result as int] >= a[i],
    }),
\end{lstlisting}
\vspace{-\baselineskip}
\begin{lstlisting}[frame=none, firstnumber=21]
{
    let mut x: usize = 0;
    let mut y: usize = a.len() - 1;
    ...
    while x != y 
\end{lstlisting}
\vspace{-\baselineskip}
\begin{lstlisting}[frame=none, firstnumber=26, backgroundcolor=\color{pink}]
        invariant
            ...
            x as int <= max_idx && max_idx <= y as int,
            forall|i: int| 0 <= i && i < a.len() ==> a[max_idx] >= a[i],
            decreases y - x
\end{lstlisting}
\vspace{-\baselineskip}
\begin{lstlisting}[frame=none, firstnumber=31]
    {
        if a[x] <= a[y] {
\end{lstlisting}
\vspace{-\baselineskip}
\begin{lstlisting}[frame=none, firstnumber=34, backgroundcolor=\color{lightcoral}]
            proof {
                ...
                if max_idx == x as int {
                    assert(a[x] == a[y]); 
                    max_idx = y as int;
                }
                ...    
                assert(x as int + 1 <= max_idx && max_idx <= y as int);
            }
\end{lstlisting}
\vspace{-\baselineskip}
\begin{lstlisting}[frame=none, firstnumber=43]
            x += 1;
        } else {
\end{lstlisting}
\vspace{-\baselineskip}
\begin{lstlisting}[frame=none, firstnumber=45, backgroundcolor=\color{lightcoral}]
            proof {
                ...
            }
 \end{lstlisting} 
\vspace{-\baselineskip}
\begin{lstlisting}[frame=none, firstnumber=48]
            y -= 1;
        }
    }
    ...
    return x;
}
\end{lstlisting}
\vspace{-\baselineskip}

    \caption{An example challenge stated in natural language highlighted in \sethlcolor{codegreen}\hl{green} and its potential solution in Verus with \texttt{code} implementation in \sethlcolor{backcolour}\hl{grey}, \texttt{spec} in \sethlcolor{codeyellow}\hl{yellow} and \texttt{proof} in \sethlcolor{lightcoral}\hl{orange} and \texttt{invariants} (a special kind of proof) in \sethlcolor{pink}\hl{pink}. This challenge is from 2011 and the solution is generated by Claude-3.7-Sonnet.}
    \label{fig:verus_code}
\end{figure}

\section{Declaration of LLM Usage}
This research evaluates LLM's performance on formal verification tasks. As for the paper preparation, LLM is ONLY used to polish the writing. 
%Bibliography
% \newpage

\end{document}